\documentclass[aps,prd,showpacs,twocolumn,superscriptaddress]{revtex4-1}
\usepackage{amsmath}
\usepackage{amsfonts}
\usepackage{graphicx}
\usepackage{dcolumn}
\usepackage{amsbsy}
\usepackage{bm}
\usepackage{amsthm}
\usepackage[caption=false]{subfig}
\usepackage{tabularx}
\usepackage{array}
\usepackage{natbib}
\usepackage{hyperref}

\begin{document}
\title{Nonthermal Spectrum of Hawking Radiation}

\author{Yu-Lei Feng}
\email{11336008@zju.edu.cn}
 \affiliation{Zhejiang Institute of Modern Physics, Zhejiang University, Hangzhou 310027, China}
\author{Yi-Xin Chen}
 \email{yixinchenzimp@zju.edu.cn}
 \affiliation{Zhejiang Institute of Modern Physics, Zhejiang University, Hangzhou 310027, China}

\begin{abstract}
  We show that for the thermal spectrum of Hawking radiation black hole's information loss paradox may still be present, even if including the entanglement information stored in the entangled Minkowski vacuum. And to avoid this inconsistency, the spectrum of Hawking radiation must be nonthermal. After reconsidering the derivation of Hawking effect, we find that the thermal spectrum is actually resulted from the geometric optics approximation in deriving the Bogolubov coefficients. When treated a little more accurately, we obtain some nonthermal spectrum for the Hawing radiation, which reduces to the thermal one in the geometric optics approximation.
\end{abstract}
\maketitle

\section{Introduction}
\label{sec:i}

Hawking effect~\cite{a,b}, which is believed to cause black hole evaporation, is a mysterious feature of quantum fields in a curved spacetime. However, the spectrum of the evaporated particles seems to be thermal, leading to the information loss paradox~\cite{b} for the black hole evaporation.
Another firewall paradox~\cite{c,d,e,f,f1,g,g1} was proposed by Almheiri et al, by means of the monogamy of entanglement. It roughly says the near horizon region will become a firewall, if black hole evaporation is unitary.
To resolve the above paradoxes, it needs to find a way to obtain pure state of Hawking radiation, meanwhile preserving the entanglement within the Minkowski (or in-falling) vacuum. In this way, quantum mechanics and general relativity may be combined consistently.

However, after a careful inspection we find that, for the thermal spectrum of Hawking radiation the information loss paradox may still be present, even though including the entanglement information stored in the entangled Minkowski vacuum, as analyzed in section~\ref{sec:ii}. And to avoid this stronger inconsistency, the spectrum of Hawking radiation must be nonthermal. Then, in section~\ref{sec:iii}, we make a detail investigation on the derivation of Hawking effect, and find that the thermal spectrum is caused by the geometric optics approximation in deriving the Bogolubov coefficients. Moreover, we obtain a nonthermal spectrum which reduces to the thermal one in the geometric optics approximation. Paralleled to the exterior mode, the interior mode is also reconsidered in section~\ref{sec:iv}. Some discussions and conclusions are given in section~\ref{sec:v} and section~\ref{sec:vi}.

\section{Information loss including entanglement information}
\label{sec:ii}

In this paper, we use the notations mainly from the reference~\cite{k2}. For a massless scalar field, the Bogolubov transformation for the s-wave modes at the infinite past $I^-$ ($a_{\omega}$) and the infinite future $I^+$ ($b_{\omega}$) is given by
\begin{equation}
\label{eq:x}
\begin{split}
b_{\omega}=\int_{0}^{\infty}d\omega'(\alpha^*_{\omega\omega'}a_{\omega'}-\beta^*_{\omega\omega'}a^{\dag}_{\omega'})
\,.
\end{split}
\end{equation}
Now, we define two operators as follows
\begin{equation}
\label{eq:y}
\begin{split}
B^1_{\omega}=\sqrt{1-e^{-8\pi M\omega}}\int_{0}^{\infty}d\omega'\alpha^*_{\omega\omega'}a_{\omega'},\\
B^2_{\omega}=\sqrt{1-e^{-8\pi M\omega}}e^{4\pi M\omega}\int_{0}^{\infty}d\omega'(-\beta_{\omega\omega'})a_{\omega'},
\end{split}
\end{equation}
which satisfy the commutation relations
\begin{equation}
\label{eq:y1}
\begin{split}
[B^i_{\omega},B^{j\dag}_{\omega'}]=\delta^{ij}\delta(\omega-\omega'),~i,j=1,2,
\end{split}
\end{equation}
where the Bogolubov coefficients for the thermal spectrum have been substituted. Notice that $[B^{1\dag}_{\omega_1},B^{2}_{\omega_2}]=0$ results in a relation
\begin{equation}
\label{eq:y2}
\begin{split}
\int_{0}^{\infty}d\omega'\alpha_{\omega_1\omega'}\beta_{\omega_2\omega'}=0.
\end{split}
\end{equation}
When calculating the correlation $\langle0_{in}\arrowvert N^{b}_{\omega_1}N^{b}_{\omega_2}\arrowvert0_{in}\rangle$ for the initial Minkowski vacuum, the relation in Eq.~\eqref{eq:y2} can lead to the \emph{uncorrelated thermal} distribution~\cite{k2}.

Now, the original $b_{\omega}$ mode can be rewritten as
\begin{equation}
\label{eq:z}
\begin{split}
b_{\omega}=\frac{B^1_{\omega}+e^{-4\pi M\omega}B^{2\dag}_{\omega}}{\sqrt{1-e^{-8\pi M\omega}}}
\,.
\end{split}
\end{equation}
Since the new $B^1_{\omega}$ and $B^2_{\omega}$ modes are independent, there should be another mode
\begin{equation}
\label{eq:z1}
\begin{split}
\tilde{b}_{\omega}=\frac{B^2_{\omega}+e^{-4\pi M\omega}B^{1\dag}_{\omega}}{\sqrt{1-e^{-8\pi M\omega}}},
\end{split}
\end{equation}
which is independent with the $b_{\omega}$ mode. Easily to see, $\tilde{b}_{\omega}$ \emph{cannot} be an \emph{exterior} mode, otherwise an apparent inconsistency will occur for the whole derivation.

The only possibility is that $\tilde{b}_{\omega}$ may be an interior mode, by choosing the interior mode function suitably. Since $B^1_{\omega}\arrowvert0_{in}\rangle=B^2_{\omega}\arrowvert0_{in}\rangle=0$, then we have
\begin{equation}
\label{eq:z3}
\begin{split}
(\tilde{b}_{\omega}-e^{-4\pi M\omega}b^{\dag}_{\omega})\arrowvert0_{in}\rangle=(b_{\omega}-e^{-4\pi M\omega}\tilde{b}^{\dag}_{\omega})\arrowvert0_{in}\rangle=0,
\end{split}
\end{equation}
leading to a relation $\arrowvert0_{in}\rangle\sim\prod_{\omega}\sum_{n}e^{-4\pi nM\omega}\arrowvert n_\omega,\tilde{n}_\omega\rangle$. Although $\arrowvert0_{in}\rangle$ is expressed as an entangled state, there seems to be no extra meaningful information stored in the entanglement. This can be seen by noting the following relation about the particle number
\begin{equation}
\label{eq:z4}
\begin{split}
\langle0_{in}\arrowvert N^{b}_{\omega_1}\tilde{N}^{\tilde{b}}_{\omega_2}\arrowvert0_{in}\rangle=\langle0_{in}\arrowvert N^{b}_{\omega_1}N^{b}_{\omega_2}\arrowvert0_{in}\rangle,
\end{split}
\end{equation}
where the relations in Eq.~\eqref{eq:z3} have been used. That is, \emph{the information stored in the entanglement is no more than the ``thermal" information stored in the exterior modes}, at least in the sense of measurement. In particular, when $\omega_1\neq\omega_2$, no correlation can be found between the exterior and interior modes. This can also be seen by noting that the reduced density matrix $\rho^{b}=tr_{\tilde{b}}\arrowvert0_{in}\rangle\langle0_{in}\arrowvert$ has the same form as a thermal one $\rho^{th}$, so that the entanglement entropy $S[\rho^{b}]$ is roughly identified with the thermal entropy $S[\rho^{th}]$. In this sense, information can be lost during the black hole evaporation, even though the entanglement information within the entangled Minkowski vacuum $\arrowvert0_{in}\rangle$ is included.

In general, it seems that the thermal spectrum of Hawking radiation is \emph{incorrect}, due to the above inconsistency which seems to be stronger than the ordinary information loss paradox. Obviously, the inconsistency is mainly resulted from the independency of the $B^1_{\omega}$ and $B^2_{\omega}$ modes, especially from the relation in Eq.~\eqref{eq:y2}.
Thus, to avoid the inconsistency, the relation Eq.~\eqref{eq:y2} must be abandoned, i.e. the spectrum of Hawking radiation must be \emph{nonthermal}~\cite{k}.

\section{Nonthermal spectrum of Hawking radiation}
\label{sec:iii}

Since the spectrum of Hawking radiation must be nonthermal, then how thermal spectrum arises? To answer this question, we need to make a detail investigation of Hawking effect. In general, we can propose a \emph{dynamical} black hole formation process, with the
metric satisfying
\begin{equation}
\label{eq:8}
\begin{split}
\eta_{\mu\nu}\stackrel{t\rightarrow-\infty}{\longleftarrow}g_{\mu\nu}(t)\stackrel{t\rightarrow+\infty}{\longrightarrow}g^B_{\mu\nu},
\end{split}
\end{equation}
with $g^B_{\mu\nu}$ a black hole's metric. Then, the initial (s-wave)
mode function satisfies the asymptotic condition near the infinite past $I^-$
\begin{equation}
\label{eq:8a}
\begin{split}
U_{\omega}(t)\stackrel{t\rightarrow-\infty}{\longrightarrow}\frac{1}{4\pi\sqrt{\omega}}\frac{e^{-i\omega v}}{r},
\end{split}
\end{equation}
where $v=t+r$; while the final exterior mode function satisfies the condition near the infinite future $I^+$
\begin{equation}
\label{eq:8b}
\begin{split}
u_{\omega}(t)\stackrel{t\rightarrow+\infty}{\longrightarrow}\frac{1}{4\pi\sqrt{\omega}}\frac{e^{-i\omega u_*}}{r},
\end{split}
\end{equation}
where $u_*=t-r_*$ with $r_*=r+2M\ln\frac{r-2M}{2M}$, similarly for the final interior mode. These
mode functions can be related via Bogolubov coefficients as follows
\begin{equation}
\label{eq:8c}
\begin{split}
u_{\omega}=\int_{0}^{\infty}d\omega'(\alpha_{\omega\omega'}U_{\omega'}+\beta_{\omega\omega'}U^*_{\omega'}).
\end{split}
\end{equation}

\begin{figure}[tbp]
\setlength{\unitlength}{1mm} \centering
\includegraphics[angle=-90,width=2.0in]{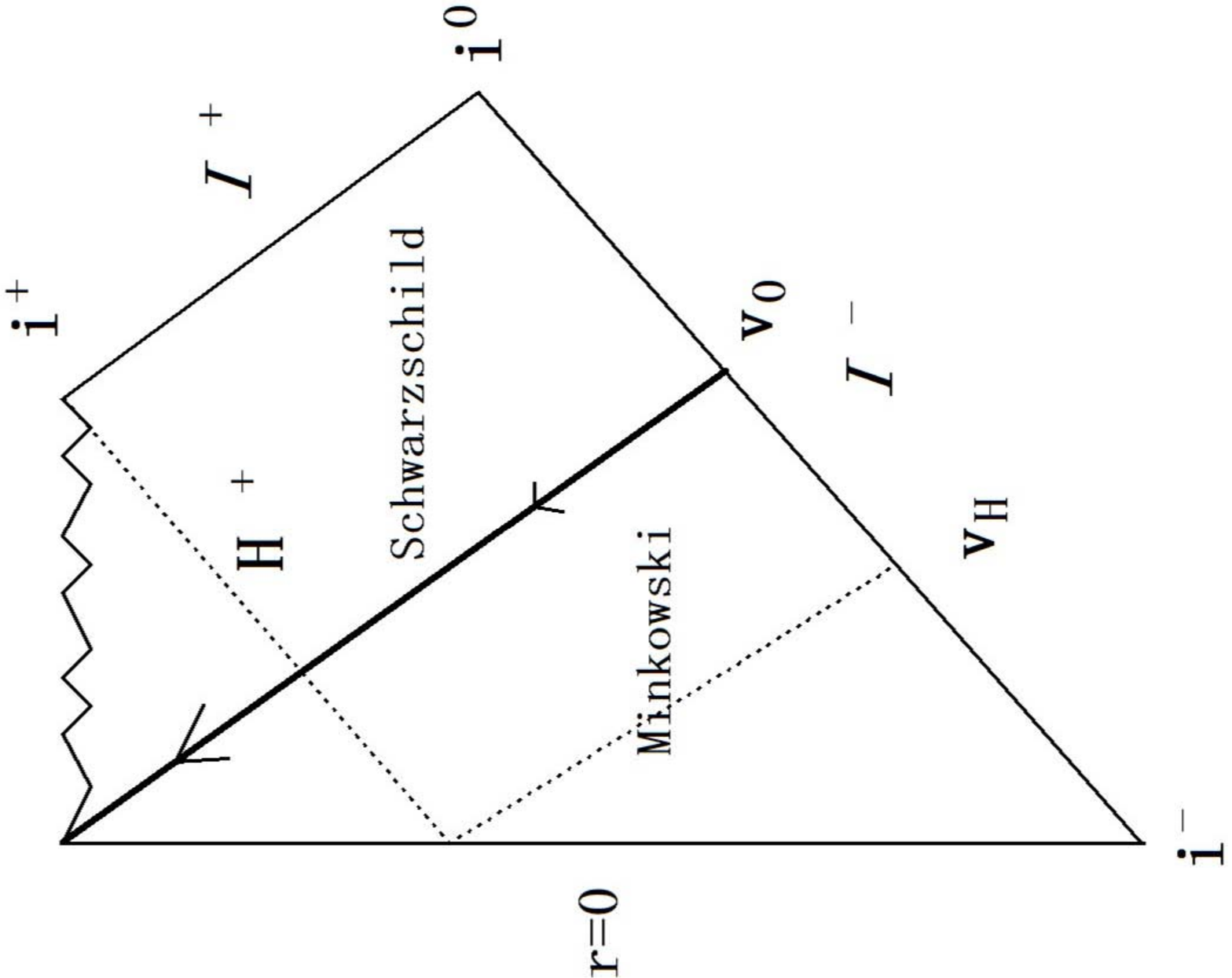}
\caption{\label{fig:1} Shock wave model. }
\end{figure}

Unfortunately, analysis for the full dynamical metric $g_{\mu\nu}(t)$ is complicated.
For simplicity, we consider a shock wave model as shown in figure~\ref{fig:1}, i.e. a Schwarzschild black hole is produced by a shock wave at $v=v_0$. The mode functions at $I^-$ and $I^+$ are given by the asymptotic form in Eqs.~\eqref{eq:8a}~\eqref{eq:8b}~\cite{k2}
\begin{equation}
\label{eq:z2}
\begin{split}
U^{as}_{\omega}=\frac{1}{4\pi\sqrt{\omega}}\frac{e^{-i\omega v}}{r},~~u^{as}_{\omega}=\frac{1}{4\pi\sqrt{\omega}}\frac{e^{-i\omega u_*}}{r},
\end{split}
\end{equation}
where superscript ``$as$" is denoted as the asymptotic form for short. After some calculations, the corresponding ``asymptotic" Bogolubov coefficients are given by
\begin{equation}
\label{eq:z5}
\begin{split}
\beta^{as}_{\omega\omega'}=\frac{-1}{2\pi}\sqrt{\frac{\omega'}{\omega}}\int_{-\infty}^{v_H}dv e^{-i\omega u_*-i\omega' v},\\
\alpha^{as}_{\omega\omega'}=\frac{-1}{2\pi}\sqrt{\frac{\omega'}{\omega}}\int_{-\infty}^{v_H}dv e^{-i\omega u_*+i\omega' v},
\end{split}
\end{equation}
which are approximations of the exact Bogolubov coefficients in Eq.~\eqref{eq:8c}
\begin{equation}
\label{eq:z5a}
\begin{split}
\beta_{\omega\omega'}\sim\beta^{as}_{\omega\omega'},~~\alpha_{\omega\omega'}\sim\alpha^{as}_{\omega\omega'}.
\end{split}
\end{equation}

Here add some notes before calculating the ``asymptotic" Bogolubov coefficients. Since ``asymptotic" Bogolubov coefficients are not exact Bogolubov coefficients, they may not satisfy the normalization condition, i.e.
\begin{equation}
\label{eq:7e}
\begin{split}
\int_{0}^{\infty}d\omega'\left[(\alpha^{as}_{\omega_1\omega'})^*\alpha^{as}_{\omega_2\omega'}-(\beta^{as}_{\omega_1\omega'})^*\beta^{as}_{\omega_2\omega'}\right]\neq\delta(\omega_1-\omega_2).
\end{split}
\end{equation}
This means a violation of the normalization condition for the exterior asymptotic mode $u^{as}_{\omega}$ at $I^-$
\begin{equation}
\label{eq:7e1}
\begin{split}
(u^{as}_{\omega_1},u^{as}_{\omega_2})|_{I^-}\neq\delta(\omega_1-\omega_2).
\end{split}
\end{equation}
However, mode $u^{as}_{\omega}$ obeys the normalization condition exactly at $I^+$. This contradiction is mainly because
that the whose spacetime is discontinuous
at $v=v_0$ for the shock wave model, i.e. $I^-$ with parameter
$v$ belongs to Minkowski spacetime while $I^+$ with
parameter $u_*$ belongs to a Schwarzschild black hole. As a result, the normalization condition is not necessarily preserved between the two sides of $v=v_0$. Certainly, this doesn't occur for the spacetime given by the metric $g_{\mu\nu}(t)$ in Eq.~\eqref{eq:8}.

To calculate the ``asymptotic" Bogolubov coefficients, it needs to find the relation between $u_*$ and $v$. It's believed that no particle is produced in the early times $u_*\rightarrow-\infty$ (or $v\rightarrow -\infty$)~\cite{k2}. In this case, $u_*\approx v$ and the ``asymptotic" Bogolubov coefficients are given by $\beta^{as}_{\omega\omega'}=0,\alpha^{as}_{\omega\omega'}=\delta(\omega-\omega')$. While the Hawking radiation comes mainly from the late times $u_*\rightarrow\infty$ (or $v\rightarrow v_H$), and in this limit we have~\cite{k2}
\begin{equation}
\label{eq:z6}
\begin{split}
u_*\approx v_H-4M\ln\frac{v_H-v}{4M}.
\end{split}
\end{equation}
By substituting Eq.~\eqref{eq:z6} into Eq.~\eqref{eq:z5}, thermal spectrum of the Hawking radiation can be obtained, with the ``asymptotic" Bogolubov coefficients satisfying $|\alpha^{as}_{\omega\omega'}|=e^{4\pi M\omega}|\beta^{as}_{\omega\omega'}|$ and the relation in Eq.~\eqref{eq:y2}. Note further that the relation Eq.~\eqref{eq:z6} has the same form as the one obtained in the geometric optics approximation~\cite{a,k2}, up to some constants. In this sense, \emph{the thermal spectrum of Hawking radiation can be considered to be mainly caused by the geometric optics approximation}.

To confirm the above statement, it still needs to show that the spectrum will not be thermal, if \emph{without} using the geometric optics approximation.
Easily to see, the above analysis by considering those two extreme cases is not accurate. Actually, the relation in Eq.~\eqref{eq:z6} is applicable only in the vicinity of $v\sim v_H$. However, the integrals in Eq.~\eqref{eq:z5} are over a large range $(-\infty,v_H)$, thus we need some more accurate relation $u_*(v)$. Note that for the shock wave model, the black hole formation is completed almost instantaneously. Thus, by means of the matching condition along the line $v=v_0$ we can obtain a more accurate relation~\cite{k2}
\begin{equation}
\label{eq:a}
\begin{split}
u_*(v)=v-4M\ln\frac{v_H-v}{4M}.
\end{split}
\end{equation}
By using of the relation Eq.~\eqref{eq:a}, some different ``asymptotic" Bogolubov coefficients can be obtained.

By substituting Eq.~\eqref{eq:a} into Eq.~\eqref{eq:z5}, we have
\begin{equation}
\label{eq:b}
\begin{split}
\beta^{as}_{\omega\omega'}=\frac{-1}{2\pi}\sqrt{\frac{\omega'}{\omega}}(4M)^{-4M\omega i}e^{-i(\omega+\omega')v_H}\\
\times\int_{0}^{\infty}dx x^{4M\omega i}e^{i(\omega+\omega')x},\\
\alpha^{as}_{\omega\omega'}=\frac{-1}{2\pi}\sqrt{\frac{\omega'}{\omega}}(4M)^{-4M\omega i}e^{-i(\omega-\omega')v_H}\\
\times\int_{0}^{\infty}dx x^{4M\omega i}e^{i(\omega-\omega')x},
\end{split}
\end{equation}
where we have define $x=v_H-v$. Further, by using of the identity
\begin{equation}
\label{eq:c}
\begin{split}
\int_{0}^{\infty}dx x^ae^{-bx}=b^{-1-a}\Gamma(1+a),
\end{split}
\end{equation}
we obtain their formal expressions
\begin{equation}
\label{eq:e}
\begin{split}
\beta^{as}_{\omega\omega'}=\frac{-1}{2\pi}\sqrt{\frac{\omega'}{\omega}}(4M)^{-4M\omega i}e^{-i(\omega+\omega')v_H}\\
\times[-i(\omega+\omega')+\epsilon]^{-1-4M\omega i}\Gamma(1+4M\omega i),\\
\alpha^{as}_{\omega\omega'}=\frac{-1}{2\pi}\sqrt{\frac{\omega'}{\omega}}(4M)^{-4M\omega i}e^{-i(\omega-\omega')v_H}\\
\times[-i(\omega-\omega')+\epsilon]^{-1-4M\omega i}\Gamma(1+4M\omega i), ~(\omega\neq\omega'),
\end{split}
\end{equation}
where an infinitesimal parameter $\epsilon$ is introduced to make the integrals to converge.

The formal expression for $\alpha^{as}_{\omega\omega'}$ is restricted by $\omega\neq\omega'$, while $\alpha^{as}_{\omega\omega}$ is divergent, i.e. it's the mode $U^{as}_{\omega}$ at $I^-$ that mainly contributes to the mode $u^{as}_{\omega}$ at $I^+$. Moreover, in the geometric optics approximation or short wavelength limit ($\omega,\omega'\rightarrow\infty$)
\begin{equation}
\label{eq:e1}
\begin{split}
|\beta^{as}_{\omega\omega'}|\approx0, ~|\alpha^{as}_{\omega\omega'}|\approx|\alpha^{as}_{\omega\omega}|,
\end{split}
\end{equation}
this roughly coincides with the classical descriptions of the light ray reflection from $r=0$ to the infinity, before the black hole is formed.

From Eq.~\eqref{eq:e}, we can see that $|\alpha^{as}_{\omega\omega'}|\neq e^{4\pi M\omega}|\beta^{as}_{\omega\omega'}|$ and $\int_{0}^{\infty}d\omega'\alpha^{as}_{\omega_1\omega'}\beta^{as}_{\omega_2\omega'}\neq0$, i.e. the spectrum of Hawking radiation is \emph{nonthermal}.
We can also calculate $|\beta^{as}_{\omega\omega'}|^2$ formally as
\begin{equation}
\label{eq:f}
\begin{split}
|\beta^{as}_{\omega\omega'}|^2=\frac{2M}{\pi}\frac{\omega'}{(\omega+\omega')^2}\frac{1}{e^{8\pi M\omega}-1},
\end{split}
\end{equation}
which gives a nonthermal spectrum formally as
\begin{equation}
\label{eq:f1}
\begin{split}
\int_{0}^{\infty}d\omega'|\beta^{as}_{\omega\omega'}|^2=\frac{2M}{\pi}\left[\ln(\frac{\Lambda+\omega}{\omega})-\frac{\Lambda}{\Lambda+\omega}\right]\frac{1}{e^{8\pi M\omega}-1},
\end{split}
\end{equation}
where $\Lambda$ is a ultraviolet cutoff. Easily to see, in the geometric optics approximation or short wavelength limit ($\omega\rightarrow\infty$), the spectrum in Eq.~\eqref{eq:f1} reduces to the thermal one up to some constant, confirming the statement below Eq.~\eqref{eq:z6}.

\section{Interior modes}
\label{sec:iv}

In standard derivations, the asymptotic interior mode tracing back in time to $I^-$ is chosen as~\cite{k2,m,n}
\begin{equation}
\label{eq:g}
\begin{split}
\tilde{u}^{as}_{\omega}=-\frac{1}{4\pi\sqrt{\omega}}\frac{e^{i\omega(v_H-4M\ln\frac{v-v_H}{4M})}}{r}\theta(v-v_H).
\end{split}
\end{equation}
One may think that similar analysis can be made simply by using of a similar accurate relation
\begin{equation}
\label{eq:7}
\begin{split}
u_*(v)=v-4M\ln\frac{v-v_H}{4M}.
\end{split}
\end{equation}
However, after substituting this relation we will obtain some \emph{strange} results, in particular
\begin{equation}
\label{eq:7a}
\begin{split}
\eta^{as}_{\omega\omega'}\sim[-i(\omega-\omega')+\epsilon]^{-1+4M\omega i}\\
\gamma^{as}_{\omega\omega'}\sim[-i(\omega+\omega')+\epsilon]^{-1+4M\omega i},
\end{split}
\end{equation}
where the relation for the exact mode functions is
\begin{equation}
\label{eq:7b}
\begin{split}
\tilde{u}_{\omega}=\int_{0}^{\infty}d\omega'(\gamma_{\omega\omega'}U_{\omega'}+\eta_{\omega\omega'}U^*_{\omega'}).
\end{split}
\end{equation}
This means that a \emph{divergent} amount of Hawking radiation can be produced in the black hole interior, due to the singular structure of $\eta^{as}_{\omega\omega'}$. Moreover, in the geometric optics approximation or short wavelength limit $|\gamma^{as}_{\omega\omega'}|\approx0$, this does not coincides with the classical descriptions of light ray confined by the event horizon, after the black hole has been formed. All these imply that the choice of asymptotic interior mode in Eq.~\eqref{eq:g} is \emph{incorrect}.

In order to obtain some reasonable results, the asymptotic interior mode can be chosen as
\begin{equation}
\label{eq:7c}
\begin{split}
\tilde{u}^{as}_{\omega}=-\frac{1}{4\pi\sqrt{\omega}}\frac{e^{-i\omega(v-4M\ln\frac{v-v_H}{4M})}}{r}\theta(v-v_H),
\end{split}
\end{equation}
where the relation in Eq.~\eqref{eq:7} has been used. In this case, the corresponding ``asymptotic" Bogolubov coefficients will be given by
\begin{equation}
\label{eq:7c1}
\begin{split}
\eta^{as}_{\omega\omega'}=\frac{-1}{2\pi}\sqrt{\frac{\omega'}{\omega}}(4M)^{-4M\omega i}e^{-i(\omega+\omega')v_H}\\
\times\int_{0}^{\infty}dx x^{4M\omega i}e^{-i(\omega+\omega')x},\\
\gamma^{as}_{\omega\omega'}=\frac{-1}{2\pi}\sqrt{\frac{\omega'}{\omega}}(4M)^{-4M\omega i}e^{-i(\omega-\omega')v_H}\\
\times\int_{0}^{\infty}dx x^{4M\omega i}e^{-i(\omega-\omega')x},
\end{split}
\end{equation}
where $x=v-v_H$. Comparing with the coefficients in Eq.~\eqref{eq:b} for the exterior mode, we can see that the results for both exterior and interior modes are similar, except the difference resulted from the sign of $v-v_H$, which comes from $r_*=r+2M\ln\frac{|r-2M|}{2M}$. Furthermore, we can also obtain the formal expressions
\begin{equation}
\label{eq:7d}
\begin{split}
\eta^{as}_{\omega\omega'}=\frac{-1}{2\pi}\sqrt{\frac{\omega'}{\omega}}(4M)^{-4M\omega i}e^{-i(\omega+\omega')v_H}\\
\times[i(\omega+\omega')+\epsilon]^{-1-4M\omega i}\Gamma(1+4M\omega i),\\
\gamma^{as}_{\omega\omega'}=\frac{-1}{2\pi}\sqrt{\frac{\omega'}{\omega}}(4M)^{-4M\omega i}e^{-i(\omega-\omega')v_H}\\
\times[i(\omega-\omega')+\epsilon]^{-1-4M\omega i}\Gamma(1+4M\omega i), ~(\omega\neq\omega').
\end{split}
\end{equation}
Again, in the geometric optics approximation or short wavelength limit ($\omega,\omega'\rightarrow\infty$) we have
\begin{equation}
\label{eq:7d1}
\begin{split}
|\eta^{as}_{\omega\omega'}|\approx0, ~|\gamma^{as}_{\omega\omega'}|\approx|\gamma^{as}_{\omega\omega}|,
\end{split}
\end{equation}
which roughly coincides with the classical descriptions of light ray confined by the event horizon, after the black hole has been formed.

\section{Discussions}
\label{sec:v}

The violation of normalization condition in Eq.~\eqref{eq:7e} may lead to a violation of commutation relations of $b_{\omega}$ modes, when $b_{\omega}$ is expressed in terms of the approximate ``asymptotic" Bogolubov coefficients in Eq.~\eqref{eq:e}. However, for expectation values of the Minkowski vacuum $\arrowvert0_{in}\rangle$ this violation is not serious, since they can be expressed in terms of the ``asymptotic" Bogolubov coefficients without using the commutation relations of $b_{\omega}$ modes.
Thus, although the ``asymptotic" Bogolubov coefficients are not exact, some general results may be deduced from them. First, the spectrum of Hawking radiation must be nonthermal, as shown in Eq.~\eqref{eq:f1}, otherwise some inconsistency will occur, as discussed previously below Eq.~\eqref{eq:z4}. And thermal spectrum appears only in the geometric optics approximation or short wavelength limit.

Second, it seems that the Minkowski vacuum $\arrowvert0_{in}\rangle$ can \emph{not} be expressed by the form $\prod_{\omega}\sum_{n}e^{-4\pi nM\omega}\arrowvert n_\omega,\tilde{n}_\omega\rangle$, since no direct relations can be found to provide that entangled form, only by using of those ``asymptotic" Bogolubov coefficients in Eqs.~\eqref{eq:e}~\eqref{eq:7d}. In fact, some properties about the form of $\arrowvert0_{in}\rangle$ can be found through the correlation functions, for example $\langle0_{in}\arrowvert N^{b}_{\omega_1}N^{b}_{\omega_2}\arrowvert0_{in}\rangle$
\begin{equation}
\label{eq:9}
\begin{split}
\langle N^{b}_{\omega_1}N^{b}_{\omega_2}\rangle-\langle N^{b}_{\omega_1}\rangle\langle N^{b}_{\omega_2}\rangle=\\
\langle b^{\dag}_{\omega_1}b_{\omega_2}\rangle\langle b_{\omega_1}b^{\dag}_{\omega_2}\rangle+
\langle b^{\dag}_{\omega_1}b^{\dag}_{\omega_2}\rangle\langle b_{\omega_1}b_{\omega_2}\rangle,
\end{split}
\end{equation}
where terms on the right-hand side of equal sign stands for some \emph{quantum correlations}, and $\int_{0}^{\infty}d\omega'\alpha_{\omega_1\omega'}\beta_{\omega_2\omega'}$ in Eq.~\eqref{eq:y2} comes from the last term. Easily to see, in the full quantum case, $\langle  N^{b}_{\omega_1}N^{b}_{\omega_2} \rangle$ is actually a four-point correlation function. While for Hawking's thermal spectrum, we have $\langle N^{b}_{\omega_1}N^{b}_{\omega_2}\rangle=\langle N^{b}_{\omega_1}\rangle\langle N^{b}_{\omega_2}\rangle~(\omega_1\neq\omega_2)$. That is, those quantum correlations may disappear in the geometric optics approximation or short wavelength limit, then the four-point correlation function reduces to an effective two-point one in the sense of measurement.

Similar analysis can be made for the other two correlation functions $\langle0_{in}\arrowvert N^{b}_{\omega_1}\tilde{N}^{\tilde{b}}_{\omega_2}\arrowvert0_{in}\rangle$ and $\langle0_{in}\arrowvert \tilde{N}^{\tilde{b}}_{\omega_1}\tilde{N}^{\tilde{b}}_{\omega_2}\arrowvert0_{in}\rangle$. From these, we can see that quantum correlations are present not only between the exterior and interior modes, but also within the exterior or interior modes themselves. In this sense, the form of the Minkowski vacuum $\arrowvert0_{in}\rangle$ will be more complicated than $\prod_{\omega}\sum_{n}e^{-4\pi nM\omega}\arrowvert n_\omega,\tilde{n}_\omega\rangle$. Moreover, since quantum correlations are present within the exterior modes, thus information cannot be lost completely, since the correlation information within the exterior modes can be acquired, as indicated by the nonthermal spectrum in Eq.~\eqref{eq:f1}.

Another result is that it seems that \emph{Hawking radiation cannot cause a black hole to evaporate}. This can roughly be seen by noting the conditions in Eqs.~\eqref{eq:e1}~\eqref{eq:7d1}, in the geometric optics approximation or short wavelength limit. That is, the (quantum) effect of particle production or Hawking radiation seems to be negligible in the classical limit. Since a real black hole is usually formed by some classical matter, it thus seems that Hawking radiation is \emph{not} efficient enough to carry the hole's mass outside, indicating that \emph{a black hole's information may not be acquired by only measuring the Minkowski vacuum}. Then, information loss may occur only in the sense of quantum measurement via average $\langle0_{in}\arrowvert T^{b}_{\mu\nu}\arrowvert0_{in}\rangle$, which can only cause some metric perturbation~\cite{k3}. In this sense, black hole's information loss paradox is resolved effectively.

One reasonable explanation for the Hawking radiation is treating it mainly as some quantum vacuum effect due to the curved spacetime background~\cite{k3}, just like the Casimir effect~\cite{l} due to some specific boundary condition. This still needs further investigations.

\section{Conclusions}
\label{sec:vi}

For the thermal spectrum of Hawking radiation, black hole's information may still be lost even if including the entanglement information within the Minkowski vacuum.  This inconsistency implies that the spectrum of Hawking radiation must be nonthermal. When treated a little more accurately, a nonthermal spectrum of Hawking radiation can indeed be obtained, which reduces to the thermal one in the geometric optics approximation.

\acknowledgments
This work is supported by the NNSF of China, Grant No. 11375150.


\begin{thebibliography}{99}
\bibitem{k} Another way to avoid the inconsitency is to abandon the conditions $0=\int_{0}^{\infty}d\omega'\alpha^*_{\omega_1\omega'}\alpha_{\omega_2\omega'}=\int_{0}^{\infty}d\omega'\beta^*_{\omega_1\omega'}\beta_{\omega_2\omega'}$ for $\omega_1\neq\omega_2$, so that the $B^1_{\omega}$ and $B^2_{\omega}$ modes are ill-defined. Smilarly, this also means that the spectrum of Hawking radiation must be nonthermal.
\end{thebibliography}
\end{document}